\documentclass[letter]{jpsj3}

\title{$^{75}$As NQR and NMR studies of superconductivity and electron correlations in iron arsenide LiFeAs}

\author{Zheng Li$^{1,}$\thanks{E-mail address: li@psun.phys.okayama-u.ac.jp},
Yosuke Ooe$^{2}$, Xian-Cheng Wang$^{1}$, Qing-Qing Liu$^{1}$, Chang-Qing Jin$^{1}$, Masanori Ichioka$^{2}$, and Guo-qing Zheng$^{1,2}$  
} \inst{Institute of Physics and Beijing National Laboratory for
Condensed Matter Physics, Chinese Academy
of sciences - Beijing 100190, China\\
Department of Physics, Okayama University - Okayama 700-8530, Japan 
}

\abst{We report the $^{75}$As-NQR and NMR studies on the iron arsenide superconductor LiFeAs with $T_{\rm c}
\sim 17$ K. The spin lattice relaxation rate, $1/T_{1}$, decreases below $T_{\rm c}$ without a coherence peak,
and can be fitted by gaps with s$^{\pm}$-wave symmetry in the presence of impurity scattering. In the normal
state, both $1/T_{1}T$ and the Knight shift decrease with decreasing temperature but become constant below $T
\leq 50 K$. Estimate of the Korringa ratio shows that the spin correlations are weaker than that in other
families of iron arsenides, which may account for the lower $T_{\rm c}$ in this material.}

\kword{NQR, NMR, iron arsenide, superconductivity}

\begin{document}
\maketitle

The discovery of superconductivity in LaFeAsO$_{1-x}$F$_{x}$\cite{YKamihara} has generated strong interest on
iron pnictide materials. Among them the most extensively studied materials are ReFeAsO$_{1-x}$F$_{x}$ (Re: rare
earth elements, so-called 1111 compound)\cite{GFChen, ZARenPr, ZARenNd, XHChen} and AFe$_{2}$As$_{2}$ (A: Ba,
Sr, Ca, so-called 122 compound)\cite{MRotter, GFChenSr, MSTorikachvili} systems. Both of their parent compounds
show a spin density wave (SDW) ordering and a tetragonal-orthorhombic structural phase transition. With doping,
SDW is suppressed and superconductivity emerges. The superconductivity is found to be in the spin-singlet
state\cite{KMatanoPr, KMatanoBa} with multiple gaps\cite{KMatanoPr, KMatanoBa, HDing}. The Fermi surfaces (FS)
consist of two hole-pockets centered at $\Gamma$ point (0, 0) and two electron-pockets around $M$ point ($\pi
,$ $\pi$)\cite{DJSing}. It has been proposed that the Fermi surface nesting between the hole- and
electron-pockets may promote spin fluctuations\cite{IIMazin, KKuroki, AVChubukov, TKariyado} and
superconductivity with s$^{\pm}$-wave symmetry\cite{IIMazin, KKuroki, FWang} that possesses isotropic gaps on
each FS with the relative phase of $\pi$ between them.

Soon after the discovery of  1111 and 122 compounds, another arsenide, LiFeAs (so-called 111
compound)\cite{XCWang, JHTapp}, was discovered to  show superconductivity even in stoichiometric composition
without SDW transition\cite{SJZhang}. It has Cu$_{2}$Sb type tetragonal structure with space group
P4/nmm\cite{JHTapp}.
The $T_{\rm c} \sim 18$ K is lower than 1111 and 122 materials. Local density approximation (LDA) shows that
there are also hole pockets on the $\Gamma$ point and electron pockets on the M point\cite{DJSing}, although
angle-resolved photoemission spectroscopy (ARPES) has not found FS nesting\cite{SVBorisenko}. A question one
can ask is why $T_{\rm c}$ is lower in this system.

In this letter, we report a study on LiFeAs ($T_{\rm c} \sim 17$ K) using the $^{75}$As nuclear
quadrupole-resonance (NQR) and $^{75}$As nuclear magnetic resonance (NMR) techniques. The measurement was
carried out down to $T/T_c<$1/10 which allows us to discuss about the gap symmetry with less ambiguity. We find
that the spin lattice relaxation rate, $1/T_{\rm 1}$, decreases below $T_{\rm c}$ without a coherence peak,
resembling the case of 1111 and 122 systems. The result can be fitted by gaps with s$^{\pm}$-wave symmetry in
the presence of impurity scattering. In the normal state, both the Knight shift and $1/T_{\rm 1}T$ decrease
with decreasing temperature. The Korringa ratio is found to be close to unity, indicating that spin
correlations are weak in LiFeAs, which may account for the lower $T_{\rm c}$ than other iron arsenides.

Two polycrystalline samples of Li$_{x}$FeAs with nominal $x=0.8$
and $x=1.1$ were prepared by employing high pressure
method\cite{XCWang}. The starting materials of Li ($99.9\%$) and
FeAs were mixed. The pellets of mixed starting materials wrapped
with gold foil were sintered at 1 GPa to 1.8 GPa, 800 ${\rm ^{o}
C}$ for 60 min. followed by quenching from high temperature before
releasing pressure. The FeAs precursors were synthesized by
sintering the mixtures of high purity Fe and As powders sealed in
an evacuated quartz tube. Powder x-ray diffraction (XRD) indicates
that the samples are of single phase. The physical properties
including the NMR results are the same. This supports that only
stoichiometric compound can be formed.\cite{JHTapp} In the
following we only present the results for the sample with nominal
$x=0.8$.

For NQR and NMR measurements, the pellets were crushed into coarse powders and sealed into epoxy (stycast)
cases. All operations were performed in a glove box filled with He gas. AC susceptibility measurements using
the NQR/NMR coils indicate that $T_{\rm c}$ for the powdered sample is 17 K at zero magnetic field and 16 K at
$\mu _{\rm 0} H=7.3$ T. NQR and NMR measurements were carried out by using a phase coherent spectrometer. The
NQR spectra were taken by changing the frequency ($\omega$) point by point, while the NMR spectra were taken by
sweeping the magnetic field at a fixed frequency. The Knight shift was determined with respect to $\omega /
\gamma$ with $\gamma =7.2919$ MHz/T. The spin-lattice relaxation rate $1/T_{1}$ was measured by using a single
saturation pulse.


\begin{figure}
\begin{center}
\includegraphics[width=5cm]{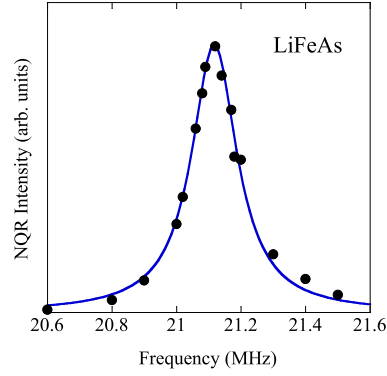}
\caption{\label{fig1} (Color online) $^{75}$As-NQR spectrum at $T=20$ K for LiFeAs. Solid curve is a Lorentzian
fitting which gives a full width at half maximum (FWHM) $\sim 0.17$ MHz.}
\end{center}
\end{figure}

Figure 1 shows the $^{75}$As-NQR spectrum at $20$ K. The clear single peak structure is observed and it can be
fitted by a single Lorentzian curve, suggesting that the sample is homogeneous. The NQR frequency $\nu _{Q} =
21.12$ MHz is much larger than LaFeAsO$_{0.92}$F$_{0.08}$ ($\nu _{Q} \sim 10.9$ MHz)\cite{SKawasaki} and
Ba$_{0.72}$K$_{0.28}$Fe$_{2}$As$_{2}$ ($\nu _{Q} \sim 5.9$ MHz)\cite{KMatanoBa}. Figure 2 shows the
$T$-dependence of $1/T_{1}$ measured by NQR at zero magnetic field and determined from an excellent fit of the
nuclear magnetization to the single exponential function $1-M(t)/M_{\rm 0}={\rm exp}(-3 t/T_{\rm 1})$, where
$M_{\rm 0}$ and $M(t)$ are the nuclear magnetization in the thermal equilibrium and at a time $t$ after the
saturating pulse, respectively. As seen in the figure, $1/T_{1}$ shows no coherence peak just below
$T_{\rm{c}}$, which is similar to other iron arsenide superconductors\cite{SKawasaki, KMatanoPr, KMatanoBa,
Nakai, Grafe}. Below $T \sim T_{\rm c}/4$, $1/T_{\rm 1}$ becomes to be proportional to $T$ in the present case,
which was also seen in BaFe$_{2}$(As$_{0.67}$P$_{0.33}$)$_{2}$ \cite{YNakai} and
LaFeAs$_{1-\delta}$O$_{0.9}$F$_{0.1}$ \cite{FHammerath}, and was explained by the existence of a residual
density of states (RDOS) in a line-node model \cite{YNakai}.

\begin{figure}
\begin{center}
\includegraphics[width=6cm]{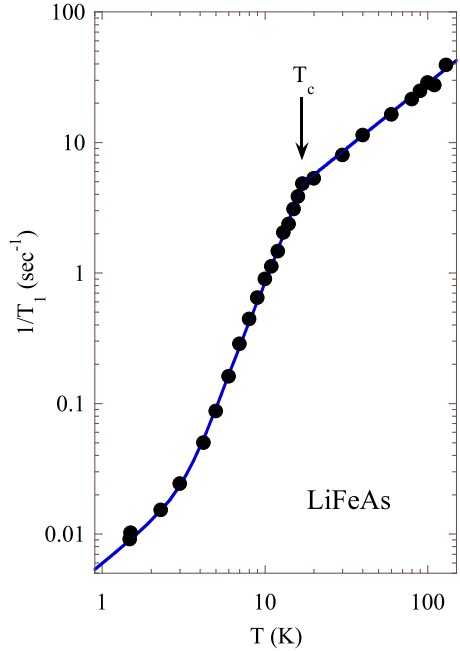}
\caption{\label{fig2} (Color online) The $T$-dependence of $1/T_{1}$ measured by NQR for LiFeAs. Experimental
error is within the size of the symbols. The curve below $T_{\rm{c}}$ is fit to the $s^{\rm \pm}$-wave model
with $\Delta_{1}^{+}=3.0$ $k_{\rm{B}} T_{\rm{c}}$, $\Delta_{2}^{-}=1.3$ $k_{\rm{B}} T_{\rm{c}}$, and the
impurity scattering rate  $\eta = 0.26$ $k_{\rm B}T_{\rm c}$ (see text).}
\end{center}
\end{figure}

Indeed, our data can also be accounted for qualitatively by a $d$-wave model\cite{KMatanoPr, KMatanoBa,
SKawasaki}, in the presence of impurity scattering. On the other hand, it has been proposed that the
$s^{\pm}$-wave is more stabilized given the Fermi surface topology.\cite{IIMazin, KKuroki} In the
$s^{\pm}$-wave model, the gaps on each FS are isotropic but the relative phase between them is $\pi$. Impurity
scattering between the electron and hole pockets could also induce a finite DOS at the Ferimi
level.\cite{YBang} However, an $s$-wave without sign change, namely $s^{++}$-wave\cite{YYanagi} would not be
able to reproduce the data at low temperature, since a finite DOS is not induced by impurity scattering in this
model. Below we show that $s^{\pm}$-wave model can fit the data quite well. By introducing the impurity
scattering parameter $\eta$ in the energy spectrum in the form of $E=\omega + i \eta$, the nuclear spin-lattice
relaxation rate $1/T_{\rm 1}$ is given by \cite{YNagai}
\begin{eqnarray}
\frac{T_{\rm 1}(T_{\rm c})}{T_{\rm 1}(T)} \cdot \frac{T_{\rm c}}{T}=\frac{1}{4T} \int _{-\infty}^{\infty}
\frac{{\rm d} \omega}{{\rm cosh}^{2} \frac{\omega}{2T}}(W_{\rm GG}+W_{\rm FF}) \label{eq:T1}.
\end{eqnarray}
where
\begin{eqnarray}
W_{\rm GG}=\left[ \left< {\rm Re} \left\{ \frac{\omega +i \eta}{\sqrt{(\omega +i \eta)^{2}+|\Delta (\bf k_{\rm F})|^{2}}} \right\} \right>_{\bf k_{\rm F}} \right] ^{2}\\
W_{\rm FF}=\left| \left< {\rm Re} \left\{ \frac{1}{\sqrt{(\omega +i \eta)^{2}+|\Delta (\bf k_{\rm F})|^{2}}}
\right\} \Delta (\bf k_{\rm F}) \right>_{\bf k_{\rm F}} \right| ^{2} \label{eq:W}.
\end{eqnarray}
Here the $\Delta$ is the gap parameter. $\left< \dots \right>$ is the average over the entire Fermi-surface,
and it sums over the contributions from two bands. Namely, for a quantity $F$,
\begin{eqnarray}
\left< F(\Delta (\bf k_{\rm F})) \right> _{\bf k_{\rm F}}=\frac{N_{1} F(\Delta _{1}) +N_{2} F(\Delta
_{2})}{N_{1} + N_{2}} \label{eq:F}.
\end{eqnarray}
Where $N_{i}$ ($i=1, 2$) is the density of state (DOS) coming from band $i$ ($i=1, 2$).  Our fitting assumes
that two gaps with different size open on different FS. It is tempting to assume that the larger gap
$\Delta_{1}$ with positive phase is on the electron FS at M point and the smaller gap $\Delta_{2}$ with
negative phase is on the hole FS at $\Gamma$ point\cite{SVBorisenko}. The parameters $\Delta_{1}^{+}=3.0$
$k_{\rm{B}} T_{\rm{c}}$, $\Delta_{2}^{-}=1.3$ $k_{\rm{B}} T_{\rm{c}}$, $N_{1} : N_{2}=0.5 : 0.5$ and $\eta =
0.26$ $k_{\rm B}T_{\rm c}$ can fit the data well as shown by the solid curve in Fig. 2. The $\Delta_{1}^{+}$ is
smaller than that in LaFeAsO$_{0.92}$F$_{0.08}$ ($\Delta_{1}=3.75$ $k_{\rm{B}} T_{\rm{c}}$, $\Delta_{2}=1.5$
$k_{\rm{B}} T_{\rm{c}}$) \cite{SKawasaki} and Ba$_{0.72}$K$_{0.28}$Fe$_{2}$As$_{2}$ ($\Delta_{1}=3.6$
$k_{\rm{B}} T_{\rm{c}}$, $\Delta_{2}=0.84$ $k_{\rm{B}} T_{\rm{c}}$) systems\cite{KMatanoBa}. The impurity
scattering parameter $\eta=\frac{\pi n_{\rm imp} (N_{1}+N_{2}) V^{2}}{1+\pi ^{2} (N_{1}+N_{2})^{2} V^{2}}$,
where $n_{\rm imp}$ is the impurity concentration and $V$ is the scattering potential at the impurity, is
slightly larger than that in LaFeAsO$_{0.92}$F$_{0.08}$ ($\eta=0.15$ $k_{\rm B}T_{\rm c}$) \cite{SKawasaki} and
Ba$_{0.72}$K$_{0.28}$Fe$_{2}$As$_{2}$ ($\eta=0.22$ $k_{\rm B} T_{\rm c}$) \cite{KMatanoBa}, which accounts for
the $T$-linear behavior at low temperature. The gap values obtained in our study is comparable to ARPES value
$\Delta_{1}^{\rm ARPES}=2.3$ $k_{\rm{B}} T_{\rm{c}}$, $\Delta_{2}^{\rm ARPES}=1.6$ $k_{\rm{B}}
T_{\rm{c}}$\cite{SVBorisenko}.

The results suggest that the  superconducting properties are quite similar to 1111 and 122 systems. This in
turn suggests that the microscopic electronic structure is similar to 1111 and 122 systems, as suggested by LDA
calculation\cite{DJSing}. In fact, if there is a large difference in the FS topology, the superconducting
property was shown to be drastically different. In  LaNiAsO$_{1-x}$F$_{x}$ ($T_{\rm{c}} \sim 4$ K), there is no
hole pocket around $\Gamma$ point\cite{GXu} and the FS nesting can not happen. There, $1/T_{\rm 1}$ shows a
well-defined coherence peak just below $T_{\rm{c}}$ followed by an exponential decay at lower
temperature\cite{TTabuchi}.

\begin{figure}
\begin{center}
\includegraphics[width=5cm]{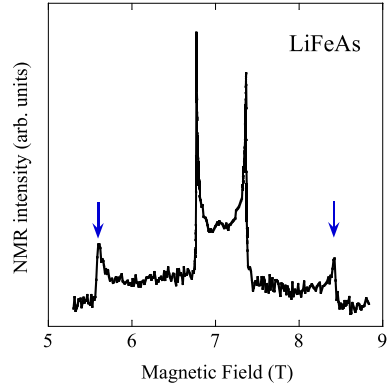}
\caption{\label{fig3} (Color online) $^{75}$As-NMR spectrum at 20 K with $f=51.1$MHz. The Knight shift and
$T_{1}$ were both measured at the left peak of the central transitions, which corresponds to $H \parallel
a$-axis.}
\end{center}
\end{figure}

In order to gain insights into the normal state properties, we also performed NMR measurements. Figure 3 shows
the $^{75}$As-NMR spectrum in LiFeAs at $T=20$ K. The two satellites indicated by the arrows correspond to the
$(\frac{1}{2} \leftrightarrow \frac{3}{2})$ and $(-\frac{3}{2} \leftrightarrow -\frac{1}{2})$ transitions. The
$\nu _{\rm Q}$ estimated from the distance between the two satellites agrees well with the NQR result. The left
peak of the central transitions (the two singularities in the center) corresponds to $H
\parallel a$-axis. Both the Knight shift and $T_{1}$ were measured at this peak. The right peak of the central
transition corresponds to $\theta = 41.8 ^{\rm o}$, where $\theta$ is the angle between $H$ and the $c$-axis.
NQR measurements correspond to $H \parallel c$-axis, since the principal axis of the NQR tensors is along the
$c$-axis.

\begin{figure}
\begin{center}
\includegraphics[width=6cm]{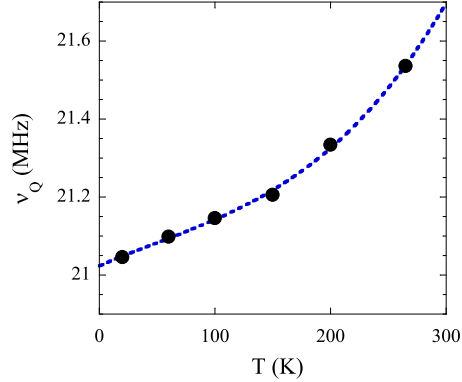}
\caption{\label{fig4} (Color online) $\nu _{\rm Q}$ as a function of temperature. The dotted curve is a guide
to the eyes.}
\end{center}
\end{figure}

\begin{figure}
\begin{center}
\includegraphics[width=7cm]{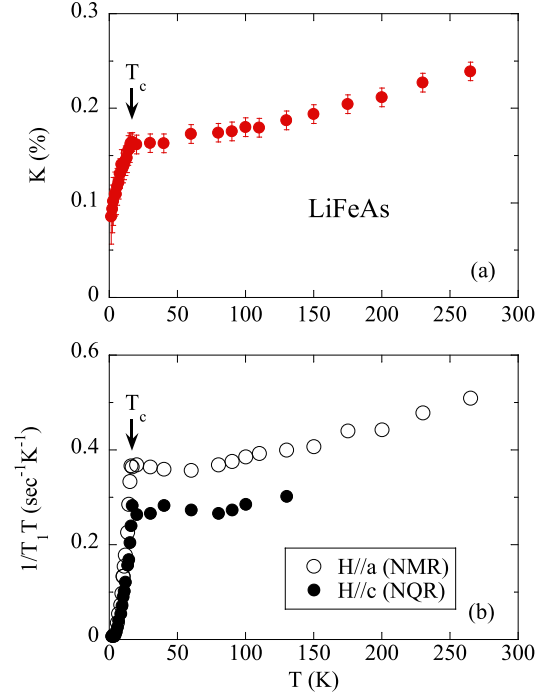}
\caption{\label{fig5} (Color online) (a) The $T$-dependence of the Knight shift. (b) The temperature dependence
of the $^{75}(1/T_{1}T)$ measured by NMR and NQR. Experimental error is within the size of the symbols.}
\end{center}
\end{figure}

The measured shift of the central transition corresponding to $H \parallel a$ consists of the Knight shift and
the shift due to second-order perturbation of the nuclear quadrupole interaction in the case of nuclear spin
$I=\frac{3}{2}$
\begin{eqnarray}
\frac{\omega-\gamma H_{res}}{\gamma H_{res}}=K+\frac{3}{16} \frac{\nu _{\rm Q }^{2}}{(1+K)(\gamma H_{res})^{2}}
\label{eq:Knight}.
\end{eqnarray}
where $H_{res}$ is the field corresponding to the peak of $H \parallel a$. In order to get accurate $K$,
$\frac{\omega-\gamma H_{res}}{\gamma H_{res}}$ vs $1/(\gamma H_{res})^{2}$ was plotted. From the linear
fitting, $K$ was deduced as the intercept, while from the slope of the plot $\nu _{Q}$ can be obtained. The
$T$-dependence of $\nu _{Q}$ so obtained is shown in Fig. 4, which decreases with decreasing temperature.
Figure 5 (a) shows the temperature dependence of $K$. It decreases with decreasing temperature but becomes
almost constant below $T=50$ K, which resembles the cases of NaCoO$_{2} \cdot$1.3H$_{2}$O
superconductor\cite{GQZheng}. Our result is consistent with a recent report by Jeglic {\it et
al}\cite{PJeglic}.

We also measured the $1/T_{\rm 1}$ corresponding to $H \parallel a$. The data are shown in the form of
$1/T_{\rm 1}T$ in Fig. 5 (b).  The temperature dependence of $1/T_{\rm 1}T$ has similar behavior as the Knight
shift. It also decreases with decreasing temperature and becomes almost constant below 50 K. The result is
different from that obtained by Jeglic {\it et al} who reported a constant $1/T_{\rm 1}T$ up to room
temperature\cite{PJeglic}. The anisotropy between $H \parallel c$ and $H \parallel a$ is consistent with that
observed in 122 system\cite{KMatanoBa}.

\begin{figure}
\begin{center}
\includegraphics[width=6cm]{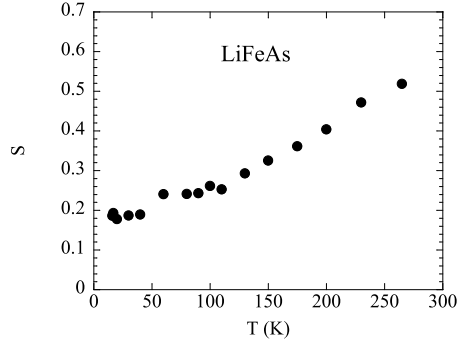}
\caption{\label{fig6} (Color online) The temperature dependence of Korringa ratio $S$.}
\end{center}
\end{figure}

The constant behavior of both $1/T_{\rm 1}T$ and $K$ below $T \leq 50$ K suggests that the electron
correlations are weak. This is in contrast to the Ba$_{0.72}$K$_{0.28}$Fe$_{2}$As$_{2}$\cite{KMatanoBa} and
LaFeAsO$_{0.92}$F$_{0.08}$\cite{SKawasaki}, where $1/T_{\rm 1}T$ increase with decreasing $T$ down to $T_{\rm
c}$. In order to evaluate more quantitatively the strength of the electron correlations, it is useful to
estimate the quantity $T_{\rm 1}TK_{\rm s}^{2}$ where $K_{\rm s}$ is the Knight shift due to spin
susceptibility. The so-called Korringa ratio $S$
\begin{eqnarray}
S=\frac{T_{\rm 1}TK_{\rm s}^{2}}{ \frac{\hbar}{4 \pi k_{\rm B}} \frac{\gamma_{e}^2}{\gamma_{n}^2}}
\label{eq:beta}.
\end{eqnarray}
where $\gamma_{e}$ and $\gamma_{n}$ are the electron and nuclear gyromagnetic ratios, is unity for
noninteracting Fermi gas. For strongly antiferromagnetically correlated metals, $S \ll 1$. For
ferromagnetically correlated metals, on the other hand, $S \gg 1$. The quantity $S$ of LiFeAs is plotted in
Fig. 6, assuming $K_{orb} \sim 0.085\%$ which is the value of $K$ at $1.5$ K. $S$ is close to 1, suggesting
that the spin correlations in LiFeAs are weak, which is in accordance with theoretical prediction\cite{ZLi}.
This conclusion does not depend on the choice of the value of $K_{orb}$. For example, if one assumes $K_{orb} =
0$, then  $S(T=265$ K$)=1.25$ and $S(T=16$ K$)=0.81$. For comparison, we note that $S$ is  1/50 in the
electron-doped cuprate superconductor due to antiferromagnetic spin correlations, even though the relation
$T_{\rm 1}TK_{\rm s}^{2}$=const. holds \cite{Zheng-2}. Such weak correlations may account for the lower $T_{\rm
c}$ in LiFeAs than other iron arsenides where electron correlations are strong or moderate. Finally, we comment
on the temperature dependence of $S$. As seen in Fig. 6, with decreasing temperature, $S$ decreases but becomes
almost constant below $T=50$ K. The origin for this $T$-dependence is unclear at present. One possibility is
that there does exist some kind of weak spin correlation at a finite wave vector $\mathbf{q}$. Since $1/T_{\rm
1}T$ is proportional to the sum of susceptibility over all  $\mathbf{q}$ but the Knight shift is decided only
by the susceptibility at $\mathbf{q}=0$, it may result in the decrease of $S$. In any case, the value of $S$
close to unity indicates that such $\mathbf{q}$-dependent correlation is not strong and it ceases to develop
below $T$=50 K.

In summary, we have performed NQR and NMR measurements on LiFeAs with $T_{\rm c} \sim 17$ K. Below $T_{\rm c}$,
$1/T_{1}$ decreases with no coherence peak and can be fitted by gaps with s$^{\pm}$-wave symmetry in the
presence of impurity scattering. The gaps obtained are $\Delta _{\rm 1}^{+}=3.0$ $k_{\rm{B}} T_{\rm{c}}$ and
$\Delta _{\rm 2}^{-}=1.3$ $k_{\rm{B}} T_{\rm{c}}$, the former is smaller than that in 1111 and 122 systems. In
the normal state, both $1/T_{\rm 1}T$ and the Knight shift decrease with decreasing temperature but becomes
constant below $T \leq 50 K$. Estimate of the Korringa ratio indicates that the electron correlations are weak
in LiFeAs, which may account for its lower $T_{\rm c}$.

We thank S. Kawasaki, K. Matano, T. Tabuchi for help in some of the measurements. This work was supported in
part by research grants from MEXT and JSPS (No. 20244058 and No. 17072005. The work at IOP was supported by
NSFC.


\end{document}